\documentclass[twocolumn]{biophys-new}
\usepackage[utf8]{inputenc}
\usepackage{graphicx}
\usepackage[colorlinks,allcolors=cyan!70!black]{hyperref}
\usepackage{amsmath}
\usepackage{amsfonts}
\usepackage{amssymb}
\usepackage{multicol}
\usepackage{caption}
\usepackage{dcolumn}
\usepackage{zref}
\usepackage{zref-xr}
\usepackage{zref-user}
\zexternaldocument{supplement}

\usepackage{lipsum}

\title{Two Heads Are (Sometimes) Better Than One: \\
  How Rate Formulations Impact Molecular Motor Dynamics}
\runningtitle{Rate Formulations and Motor Dynamics} 

\author[1,*]{R. Blackwell}
\author[1]{D. Jung}
\author[1]{M. Bukenberger}
\author[1]{A.-S. Smith}
\runningauthor{R. Blackwell} 

\affil[1]{Physics Underlying Life Sciences (PULS) group\\
  Cluster of Excellence Engineering of Advanced Materials (EAM)\\
  Department of Theoretical Physics I,  Friedrich-Alexander University, Erlangen}

\corrauthor[*]{robert.blackwell@fau.de}

\papertype{Article}

\newcommand{\mr}[1]{\mathrm{#1}}
\newcolumntype{M}[1]{>{\centering\arraybackslash}m{#1}}
\newcolumntype{N}{@{}m{0pt}@{}}

\begin{document}

\begin{frontmatter}

\begin{abstract}%
Cells are complex structures which require considerable amounts of organization via transport of large intracellular cargo.
While passive diffusion is often sufficiently fast for the transport of smaller cargo, active transport is necessary to organize large structures on short timescales.
The main mechanism of this transport is by cargo attachment to motors which walk in a directed fashion along intracellular filaments.
There are a number of models which seek to describe the motion of motors with attached cargo, from detailed microscopic to coarse phenomenological descriptions.
We focus on the intermediate-detailed discrete stochastic hopping models, and explore how cargo transport changes depending on the number of motors, motor interaction, system constraints and rate formulations which are derived from common thermodynamic assumptions.
We find that, despite obeying the same detailed balance constraint, the choice of rate formulation considerably affects the characteristics of the overall motion of the system, with one rate formulation exhibiting novel behavior of loaded motor groups moving faster than a single unloaded motor.
\end{abstract}

\end{frontmatter}

\section{INTRODUCTION}
Motor proteins are ubiquitous agents of active motion in biological systems, powering such diverse processes as the flagellar propulsion of sperm cells~\cite{ryu2000}, the contraction of muscle fibres via myosin~\cite{spudich1971} or the transport of neurotransmitters by kinesin~\cite{setou2000}. 
While studied for a multitude of reasons, two stand out for why they are of considerable research interest.
First, since motor proteins play an important role in such a wide variety of biological processes, it is not surprising that disturbances in the function of these molecular motors can cause an equally wide variety of medical problems, such as primary ciliary dyskinesia ~\cite{zariwala2007}, Bardet-Biedel syndrome~\cite{zaghloul2009}, Charcot-Marie-Tooth disease~\cite{zhao2001}, motor neuron disease~\cite{puls2003}, Alzheimer's\cite{ebneth1998}, ALS~\cite{boillee2006}, and others.
Second, they serve as an inspirational blueprint in the nascent engineering and development of synthetic molecular machines~\cite{kay2007,hanggi2009,zhang2011}.
Such synthetic motors would not only be useful due to their microscopic size per se, but would also promise an energy efficiency noticeably larger than their macroscopic counterparts.
For example, the ubiquitous kinesin motor effectively uses approximately 60\% of the chemical energy gained from hydrolyzing a single adenosine triphosphate (ATP) molecule for linear motion~\cite{bustamante2007}.
Due to these reasons, there has been a fair amount of work on modeling the motion of molecular motors, with modeling techniques falling into two main camps: collective effects of multiple motors along a single track~\cite{klumpp2005,berger2015,borodin2007,parmeggiani2004}, and the effects of the cargo on single motor dynamics~\cite{zimmermann2012,zimmermann2015,zimmermann2015a,schmiedl2008}. 

The bulk of the modeling of molecular motors has been produced on the topic of cooperative transport by groups of motors~\cite{klumpp2005,berger2015,borodin2007,parmeggiani2004}, which has been confirmed experimentally to occur in numerous in vivo systems~\cite{shubeita2008,mallik2009}.
Efforts to model cooperative transport involving direct motor-motor interactions so far have been either based on the strongly simplified ASEP (asymmetric simple exclusion process) approach, which typically use constant, load-independent jump rates~\cite{borodin2007,parmeggiani2004}, or are focused on emergent effects stemming from variable motor numbers via detachment and attachment of motors on the motor track~\cite{bouzat2010,chai2009}.

For the effects of single-motor dynamics on cargo dynamics, much research has been focused on stochastic motor hopping models, where the motor makes discrete jumps along a track with thermodynamic constraints on that motion~\cite{zimmermann2012,zimmermann2015,zimmermann2015a,schmiedl2008}.
The main constraint is one of {\itshape local detailed balance}, which strictly defines the {\itshape ratio} of the forward and backward jumping rates.
However, this constraint is insufficient to generate a unique rate formulation, forcing a choice.
We identify three of the popular model choices in the literature, which we label: Glauber~\cite{glauber1963,einax2010}, Product Asymmetric-Exchange (P-AsEx)~\cite{bhat2016,fisher1999,stukalin2005,stukalin2006}, and Difference Asymmetric-Exchange (D-AsEx)~\cite{zimmermann2012,zimmermann2015,zimmermann2015a}.

In this work we explore cargo transport properties within these models under controlled constraints to illuminate the different {\itshape physical} consequences of model choice for both single and multiple motor driven transport.
Due to the large parameter space, a complete comparative study of the different models is impractical.
Single-motor force-velocity relationships are first examined to demonstrate the varying response of the different motor models under load.
We then focus on motor-number scaling behavior of the models near physical parameters which are both experimentally accessible, and still provide a strong response to parameter perturbations while allowing for direct comparison between crowded and non-crowded motor teams.
We find that in the special case of a two-motor system that evolves with the D-AsEx rate formulation, the results are quite novel, allowing for the motor team to move faster than either motor alone, and so we pay special attention to understand the underlying physics driving the unique behavior.
In other cases, many differences are of more easily probed nature, and could be observed directly using existing experimental techniques.
We attempt to underscore these differences and propose potential experiments to explore the behavior of analogous systems to determine which model most accurately describes motor-cargo transport.

\section{METHODS}

\begin{figure}
\begin{center}
\includegraphics[width=0.45\textwidth]{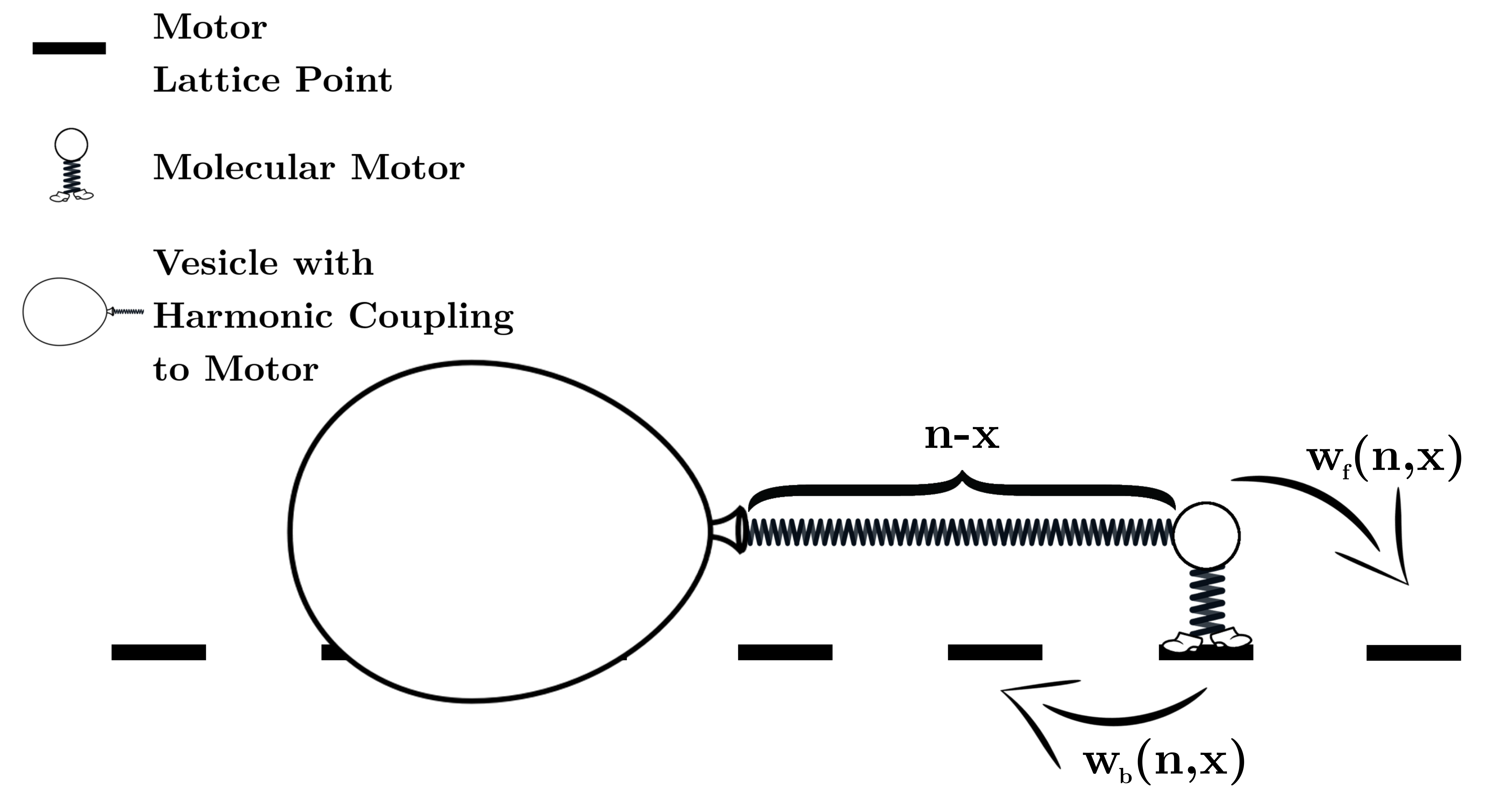}
\caption{Sketch of modeled one-dimensional motor-cargo system with harmonic coupling showing a single motor moving on a discrete lattice of positions $n$ and the cargo moving continuously along a trajectory $x(t)$.}
\end{center}
\end{figure}

Cargo transport by motor proteins takes place in a strongly viscous environment where the cargo's inertia is often negligible.
For example, in the extreme case, a large spherical cargo vesicle of radius $R=1\upmu$m with transport velocity of  $v\approx 1\upmu$m s$^{-1}$~\cite{schnitzer2000} in aqueous solution only has a Reynolds number of $\mathrm{Re}\approx 10^{-6}$.
Accounting for the fact that peak velocities of the motor resulting from its stepwise motion may be two orders of magnitude larger~\cite{holzwarth2002} we still remain in a clearly overdamped regime at $\mathrm{Re}\leq 10^{-4}$.
For this reason, it is appropriate to describe the dynamics of the one-dimensional cargo position $x(t)$ using Brownian dynamics
\begin{subequations}
    \begin{equation}\label{eq:browndyn}
      m\ddot{x}\to 0=F\left(\mathbf{n},x\right)-F_L-\gamma\dot{x}+\xi(t)\\
    \end{equation}
    \begin{equation}
      \langle \xi(t) \rangle=0\\
    \end{equation}
    \begin{equation}
    \langle \xi(t) \xi(t') \rangle= \frac{2\gamma}{\beta} \delta(t-t'),
    \end{equation}
\end{subequations}
where $F(\mathbf{n},x)$ is the systemic force on the cargo due to the motors located at sites $\mathbf{n}$ and cargo at position $x$, $F_L$ is a constant external force, $\gamma$ is the cargo's drag coefficient, the stochastic force $\xi(t)$ describes Brownian motion at a thermal energy of $1/\beta=k_{B}T$, and the cargo mass $m$ is taken to zero as inertial forces are considerably smaller than the systemic and drag forces in the overdamped limit.
In this formulation, only the particle's size affects the dynamics of transport via the Stokesian damping constant $\gamma$, which is proportional to the product of $R$ and the dynamic viscosity of the environment.
The driving force $F(\mathbf{n},x)$ results from the coupling of the cargo to the motor heads at positions $\mathbf{n}$ on the microtubule via each motor's coiled coil.
We model the stretching of this coiled coil upon motor stepping with a harmonic spring force of coupling constant $k$, whose typical stiffness is approximately on the order of a pN nm$^{-1}$~\cite{bhat2016}, or in the dimensionless microscopic units from Table~\ref{tab:units}, roughly $0.1$--$1$.

\begin{table}
  \centering
  \begin{tabular}{|M{1.0cm}|M{1.6cm}|M{1cm}|M{1.2cm}|M{1.2cm}|N}
    \hline
    \multicolumn{3}{|c|}{Base} & \multicolumn{2}{c|}{Derived} \\[2.5pt] \hline
    Length & Energy & Time & $k$ & $\gamma$ & \\[2.5pt] \hline
    $\uplambda$ &  $k_B T$ & $\tau_0$ & $\frac{k_{B}T}{\uplambda^2}$ & $\frac{k_{B}T\tau_0}{\uplambda^2}$ & \\[5pt] \hline
    8 nm & $4.3\cdot 10^{-21}$J & 1 s & $ 67\frac{\mathrm{pN}}{\upmu \mathrm{m}}$ & $67 \frac{\mathrm{pNs}}{\upmu \mathrm{m}}$ & \\[5pt] \hline
  \end{tabular}
  \caption{Our system of units used throughout this paper.}
\label{tab:units}
\end{table}

For motor dynamics, we follow prior work with a basic outline given here~\cite{zimmermann2012,zimmermann2015,zimmermann2015a,schmiedl2008}.
We first assume that motors attached to a static cargo will adopt a position probability distribution $p(n,x)$ following Maxwell-Boltzmann statistics in the long-term.
With the additional constraint that the forward and backward jump rates $w_\mr{f}$ and $w_\mr{b}$ fulfill detailed balance, we get the following central condition for our rate formulations:
\begin{equation}\label{eq:detbal1}
\frac{w_\mr{f}(n-1,x)}{w_\mr{b}(n,x)}=\frac{p(n,x)}{p(n-1,x)}=e^{-\beta\big(E(n,x)-E(n-1,x)\big)}.
\end{equation}
The state energy $E(n,x)$ can be defined iteratively by the chemical energy $\Delta \mu>0$ gained through ATP hydrolysis and the coiled coil stretching corresponding to a step of length $\uplambda$ by the motor
\begin{equation}
E(n,x)-E(n-1,x)=k\uplambda^2 (n-x-\frac{1}{2})-\Delta \mu.
\end{equation}

\subsection{Simulation}
Now that we have dynamical evolution equations for both the motor and the cargo, we can use an iterative hybrid technique to calculate the average cargo velocity by alternating between a kinetic Monte Carlo step for motor motion, and a Brownian dynamics step for the cargo motion.

For each timestep, the kinetic Monte Carlo routine is executed as follows. We first calculate the rates for each motor $i$ at position $n_i$ and cargo position $x$, to move either forward $w_{f,i}(n_i,x)$ or backward $w_{b,i}(n_i,x)$ using the relevant formulation described below.
The probability for {\itshape any} motor to move during the timestep $\delta t$ is then calculated as
\begin{equation}
  P_{\rm hop} \approx \delta t \ w_{\rm tot} = \delta t \sum_i^N \left[w_{f,i}(n_i,x)+w_{b,i}(n_i,x)\right],
\end{equation}
where $\delta t = \min(\frac{10^{-3}}{ w_{\rm tot}}, \frac{\tau}{10})$, $\tau = \frac{\gamma}{N k}$ is the cargo relaxation timescale, and $N$ is the number of motors attached to the cargo.
This constraint ensures both that multiple motor events do not occur during a given timestep, and that the cargo does not appreciably move during this timestep.
For simulations where motor heads cannot occupy the same lattice site, the appropriate rates are set to zero if the motor move would result in an overlap of two motor heads.
If an event occurs during this timestep, $\mathcal{U} < \delta t\ w_{\rm tot}$, where $\mathcal{U}$ is a uniform number on the interval $[0,1)$, then a single motor move is executed randomly with a weighted probability $P_{i,j} = w_{i,j}/w_{\rm tot}$, where $i$ represents the individual motor, and $j$ represents the direction of hopping.

After completion of the kinetic Monte Carlo portion of the iteration, the cargo then undergoes a simple 1 dimensional Brownian dynamics step,
  \begin{equation}
    x_{t+\delta t} = x_t + \frac{k\delta t}{\gamma} \sum_i^N\left(n_i-x_t\right)
    - \frac{F_L \delta t}{\gamma} + \sqrt{2~ k_B T\ \delta t/\gamma}~\mathcal{N},\label{eqn:simulation}
  \end{equation}
where $\mathcal{N}$ is a normally distributed random number with zero mean and a variance of one.

However, for the kinetic Monte Carlo step, there are a number of different formulations for the rates $w_f$ and $w_b$ which obey detailed balance, each with different dynamics depending on the input parameters (Fig. \ref{fig:singlemotor}A-C).
We outline three popular rate formulations here, with alterations where necessary.

\subsection{Rate formulations}
\subsubsection{Glauber}
The Glauber rates follow an ansatz which strongly resembles the Fermi-Dirac distribution and also fulfills the detailed balance condition of Equation~\ref{eq:detbal1}~\cite{glauber1963,einax2010}.
\begin{subequations}
\begin{equation}
w_\mr{f}^\mr{G}\left(n,x\right)=\frac{2w_0}{1+e^{k\left(n-x+\frac{1}{2}\right)}}
\end{equation}\label{eq:glauber-forward}
\begin{equation}
w_\mr{b}^\mr{G}\left(n,x\right)=\frac{2w_0\cdot e^{-\Delta \mu}}{1+e^{-k \left(n-x-\frac{1}{2} \right)}}.
\end{equation}
\end{subequations}
While not extensively studied in the literature, the Glauber rates have an advantage their simplicity: they do not introduce an additional parameter into the rate formulation.

\begin{figure*}
  \begin{center}
    \includegraphics[]{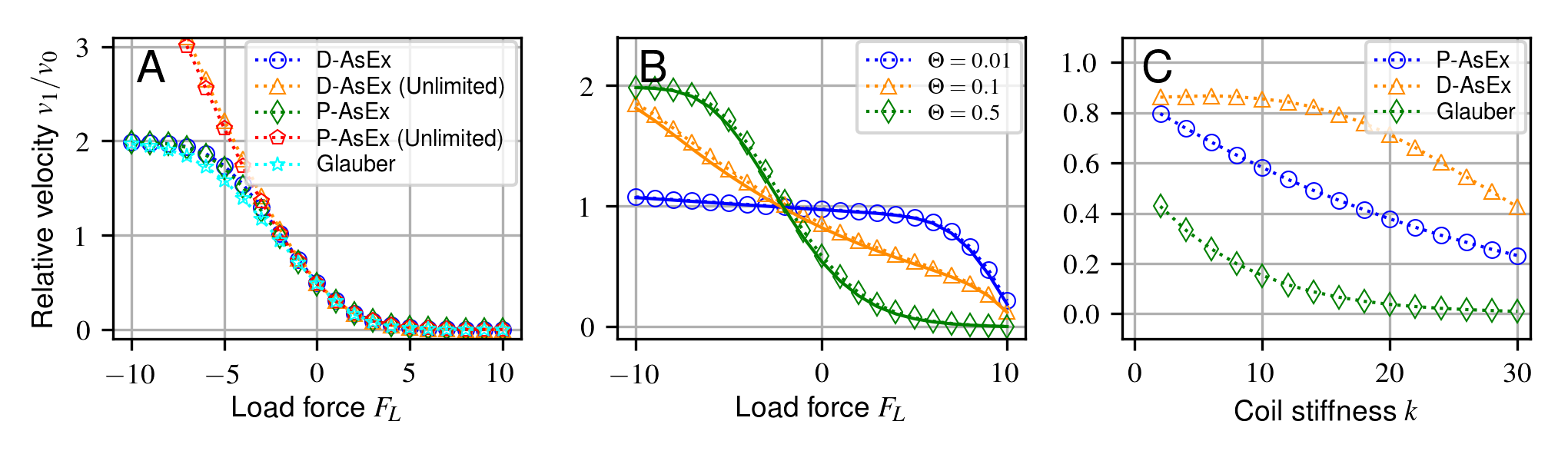}\caption{
      Transport velocities of a cargo pulled by a single motor.
      System parameters are chosen to provide an unloaded motor velocity of $v_0 = 1 \upmu$m s$^{-1}$.
      Default simulation parameters in all three plots in our dimensionless units are: $w_0=125.0028$, $\gamma=1.41\cdot 10^{-2}$, $k=1$, and $\Delta\mu=10.7065$.
      \textbf{\textit{(A)}} Force-velocity curve for each rate formulation generated by subtracting a given load force from the force term $F(\mathbf{n},x)$ in Eq.\eqref{eq:browndyn} acting on the cargo. This convention means that a positive load force {\itshape opposes} the forward motion of the motor.
      $\Theta^\mr{D}=0.76$, $\Theta^\mr{P}=0.71$ chosen to ensure equal transport velocities at zero load force.
      \textbf{\textit{(B)}} Change of force-velocity curve shape from convex to concave depending on $\Theta$ demonstrated for D-AsEx (dotted) and P-AsEx (solid) rates.
      \textbf{\textit{(C)}} Transport velocities for each rate formulation using identical parameters and varying the coiled coil stiffness $k$ with $\Theta=0.1$ in the three models.
    }\label{fig:singlemotor}
  \end{center}
\end{figure*}

A more common alternative to the Glauber ansatz is one with a formulation reminiscent of the Arrhenius equation for chemical reaction rates~\cite{bhat2016,fisher1999,zimmermann2012,zimmermann2015,zimmermann2015a,stukalin2005,stukalin2006}.
In this work, we classify this family of reaction rates as asymmetric exponential (AsEx) rates and further consider two variations of this ansatz.
In both of the AsEx rate formulations considered here, there is a free parameter $\Theta\in \mathbb{R}\cap [0,1]$ which functions as a weighting factor for the relative impact of forces acting via the coiled coil on the forwards versus the backwards rates.
In contrast to the previously studied models, however, we present these models in an amended form which removes the unbounded motor velocities under load (Fig. \ref{fig:singlemotor}A).
This amendment is necessary to agree with the finding that motor velocities saturate under forward load~\cite{block2003}.
We have therefore adjusted the forward and backward rates to never exceed some maximum rate $\alpha w_0$, where $\alpha\in \mathbb{R}\cap [0,\infty]$, by introducing Heaviside step functions $\mathcal{H}$ into the rate formulations in a way that maintains the detailed balance constraint.
In this work, we will however focus only on the $\alpha = 2$ case for direct comparison to the Glauber rates.
We distinguish two variations below, letting $\delta x = n - x$ represent the displacement between the motor and the cargo.

\subsubsection{Product Asymmetric Exchange (P-AsEx)}
Here, the parameter $\Theta$ splits the change in harmonic potential energy $V(\delta x)=\frac{1}{2} k~\delta x^2$, corresponding to a forward or backward jump respectively by factors of $\Theta$ and $(1-\Theta)$~\cite{bhat2016,fisher1999,stukalin2005,stukalin2006}.
Letting the change in effective potential for a jump from site $n \rightarrow n+1$ be denoted as $\Delta V_P\left(\delta x, \Theta\right) = \frac{k\Theta}{2} \left(\left(\delta x + 1\right)^2 - \delta x^2\right) = k \Theta\left(\delta x + \frac{1}{2}\right)$, the rates can be written
\begin{subequations}
  \begin{align}
    w_f^\mr{P}& =~w_0 e^{-\Delta V_P\left(\delta x, \Theta \right)} \times \\
                &e^{\mathcal{H}\left(-\Delta V_P\left(\delta x, \Theta\right) - \ln \alpha\right)\cdot
                  \left(\Delta V_P\left(\delta x, \Theta\right) + \ln \alpha\right)} \times \notag\\
                &e^{\mathcal{H}
                  \left( -\Delta V_P \left(\delta x + 1, \Theta - 1\right) - \ln \alpha - \Delta \mu \right)\cdot
                  \left(\Delta V_P\left(\delta x + 1, \Theta - 1\right) + \ln \alpha + \Delta \mu\right)}\notag
  \end{align}
  \begin{align}
    w_b^\mr{P}& =~w_0 e^{-\Delta \mu -\Delta V_P\left(\delta x - 1, \Theta - 1 \right)} \times \\
                &e^{\mathcal{H}\left(-\Delta V_P\left(\delta x - 1, \Theta\right) - \ln \alpha\right)\cdot
                  \left(\Delta V_P\left(\delta x - 1, \Theta\right) + \ln \alpha\right)} \times \notag\\
                &e^{\mathcal{H}\left(-\Delta V_P\left(\delta x, \Theta - 1\right) - \ln \alpha - \Delta \mu\right)\cdot
                  \left(\Delta V_P\left(\delta x, \Theta - 1\right) + \ln \alpha + \Delta \mu \right)},\notag
  \end{align}
\end{subequations}
where in each expression the first term represents the unlimited rate from prior work, the second limits the forward rates to $w_0 \alpha$, and the third limits the backward rates to $w_0 \alpha$.
The existence of a forward (backward) limiting term in the backward (forward) rate is required to maintain detailed balance.

\subsubsection{Displacement Asymmetric Exchange (D-AsEx)}
Here, $\Theta$ splits the stepping distance $\uplambda$ into substeps determining the forward and backward rates respectively of size $\Theta\lambda$ and $(1-\Theta)\lambda$.
Letting the effective potential $\Delta V_D\left(\delta x, \Theta\right) = \frac{k}{2}\left(\left(\delta x + \Theta\right)^2 - \delta x^2\right) = k \Theta\left(\delta x + \frac{\Theta}{2}\right)$, the limited rates are then
\begin{subequations}
  \begin{align}
    w_f^\mr{D}& =~w_0 e^{-\Delta V_D\left(\delta x, \Theta \right)} \times \\
                &e^{\mathcal{H}\left(-\Delta V_D\left(\delta x, \Theta\right) - \ln \alpha\right)\cdot
                  \left(\Delta V_D\left(\delta x, \Theta\right) + \ln \alpha\right)} \times \notag\\
                &e^{\mathcal{H}\left(-\Delta V_D\left(\delta x + 1, \Theta - 1\right) - \ln \alpha - \Delta \mu\right)\cdot
                  \left(\Delta V_D\left(\delta x + 1, \Theta - 1\right) + \ln \alpha + \Delta \mu\right)}\notag
  \end{align}\label{eq:d-asex-forward}
  \begin{align}
    w_b^\mr{D}& =~w_0 e^{-\Delta \mu -\Delta V_D\left(\delta x, \Theta - 1 \right)} \times \\
                &e^{\mathcal{H}\left(-\Delta V_D\left(\delta x - 1, \Theta\right) - \ln \alpha\right)\cdot
                  \left(\Delta V_D\left(\delta x - 1, \Theta\right) + \ln \alpha\right)} \times \notag\\
                &e^{\mathcal{H}\left(-\Delta V_D\left(\delta x, \Theta - 1\right) - \ln \alpha - \Delta \mu\right)\cdot
                  \left(\Delta V_D\left(\delta x, \Theta - 1\right) + \ln \alpha + \Delta \mu\right)},\notag
  \end{align}
\end{subequations}
where, as with the P-AsEx model, the first term represents the unlimited rate, the second limits the forward rates, and the third limits the backward rates.
Note that while the rates are nearly identical to the P-AsEx model with a change in the $\Delta V$ term, the backward rate in the D-AsEx model has a slight change in form compared to that of the P-AsEx model.

The split employed in the D-AsEx rates can be interpreted to imply Brownian Ratchet dynamics in an asymmetric potential between motor lattice points.
This specific rate formulation has been used in particularly by Zimmermann \& Seifert~\cite{zimmermann2012,zimmermann2015,zimmermann2015a}, and can be considered an attempt to unify the discrete stochastic model of motor dynamics with the fundamental idea of a continuum ratchet model~\cite{kolomeisky2007}.

We also note that the leading term in the D-AsEx and P-AsEx model can be mapped to between one another via a change in the effective rate constant  $w^\mr{P}_0=w^\mr{D}_0e^{-\frac{k\Theta}{2}(1-\Theta)}$.
Since $k$ in a physical system is $\sim 0.1-1$, this correction is typically quite small.
We therefore focus mostly on the D-AsEx model in this paper, with most of the corresponding figures for the P-AsEx model given in the supplement.

\section{RESULTS}
\subsection{Single Motor}
Despite sharing the same detailed balance condition, there are considerable qualitative differences between the models, even for a single motor, depending on the parameter choices.
One of the most common ways to experimentally examine motor dynamics is through experiments where a constant load force $F_L$ is supplied to the cargo via an optical trap, with force-velocity curves varying in shape from roughly linear to convex~\cite{wang1998,svoboda1994,schnitzer2000}.
We therefore first examine the cargo velocity under a constant load force to directly compare the models in Figs.~\ref{fig:singlemotor}A and \ref{fig:singlemotor}B, normalizing by the unloaded motor velocity
\begin{equation}
v_0=w_0\cdot\left(1-e^{-\beta \Delta \mu}\right)
\end{equation}
for ease of comparison between the different models.

In Fig.~\ref{fig:singlemotor}A we use the parameter $\Theta$ to equalize the transport velocities for all three rates at zero external force and otherwise identical system parameters.
All of the rate formulations show nearly identical force-velocity relationships when the force is positive, i.e.~the forces are applied against the motor's preferred direction of motion.
For forces in the motor's preferred walking direction, we find that the velocities under the Glauber rates and our modified AsEx rates all saturate at a maximum velocity given by the ATP concentration, as we would expect in a real system, whereas the original unlimited AsEx forward rates are inherently unbounded (Fig. \ref{fig:singlemotor}A). 

Despite the nearly identical behavior shown between the models when fixing $\Theta$ to high values as in Fig.~\ref{fig:singlemotor}A, the AsEx models exhibit considerably different behavior when the load-splitting parameter $\Theta$ is allowed to vary.
In both of the AsEx models, varying $\Theta$ can dramatically change the inflection and slope of the force-velocity curve near the zero-force limit~(Fig.~\ref{fig:singlemotor}B).
We also see from Figure~\ref{fig:singlemotor}B the near equivalence of the two AsEx models at low $k$, when normalized by the unloaded velocity $v_o$.
Similar inflection and slope differences have been observed in prior experiments, potentially being driven by an underlying AsEx model~\cite{wang1998,svoboda1994,schnitzer2000}.
These shape changes can be easily understood by looking at the $\Theta$ dependence in the AsEx models with $w_f \sim w_0 e^{-k\Theta\delta x}$: a small $\Theta$ leads to a low sensitivity of the forward rate in motor extension, leading to a small response to load force.

The models can also exhibit different qualitative behavior under other parameter changes.
For example, by varying the coil stiffness $k$, as shown in Fig.~\ref{fig:singlemotor}C, cargo transport velocities can vary considerably between the three rate formulations for identical parameters.
While the D-AsEx model has low sensitivity on the coil stiffness for this choice of parameters, the P-AsEx and Glauber models result in an order of magnitude decrease in the cargo velocity and a completely different functional form (Fig.~\ref{fig:singlemotor}C).

It is clear that the models exhibit distinct changes in response under load.
In principle, the response to load with known $w_0$, $\Delta \mu$, $k$, $\gamma$, and saturation velocity $w_0 \alpha$ should be sufficient to uniquely identify if the Glauber or an AsEx model is appropriate.
Due to the difficulty of adjusting the coil stiffness of the motor protein $k$, it is potentially difficult to distinguish between the two AsEx models in biological systems, since the differences at biological $k$ are small (Fig.~\ref{fig:singlemotor}).
Therefore, multiple-motor experiments may provide a more reliable way to experimentally determine the underlying motor mechanics over single-motor force-velocity measurements, as well as illuminating how motors can cooperate under various constraints.

\subsection{Two Motors}
\begin{figure}[hbt!]
  \centering
  \includegraphics[]{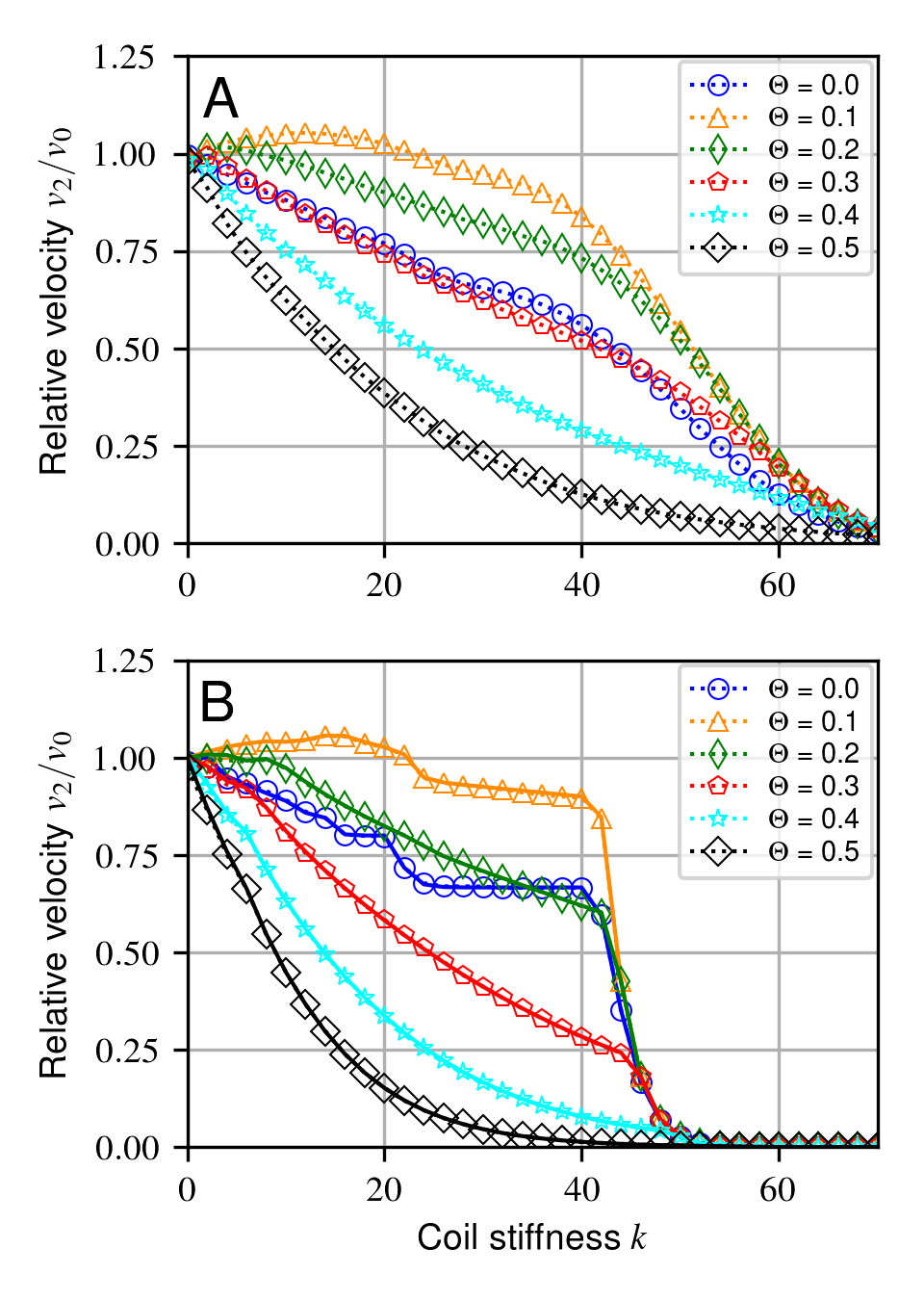}
  \caption{Relative velocities $v_1/v_0$ for two non-excluding motors and with parameters in our dimensionless units $\gamma=10^{-3}$,$\Delta\mu=20$, and $w_0=60$.
    \textbf{\textit{(A)}} D-AsEx rates simulation for two non-interacting motors.
    \textbf{\textit{(B)}} D-AsEx rates simulation (dotted line) vs analytic approximation (full line) with no thermal motion of the cargo.
  } \label{fig:fasterthanfree}
\end{figure}
When examining cooperativity of motors operating in teams, we discovered a novel behavior in systems of two motors in certain limits: two non-interacting {\itshape loaded} motors can move faster than each individual {\itshape unloaded} motor can alone in the D-AsEx model, with the effect being most prominent for relatively large values of the motor coil stiffness $k$ (Fig. \ref{fig:fasterthanfree}A).
This effect appears in the low drag ($\gamma \ll 1$), and high biasing ($\Delta \mu \sim 10-20$) limit, with the effect being most prominent at large coil stiffness ($k\sim 10-20$).
This regime might difficult explore in biological systems, due to the need to control the motor intrinsic coil stiffness $k$, potentially by direct mutations or interactions with other molecules or proteins.
However, the findings here could be readily useful for engineered motors, where the coil stiffness could potentially be exploited directly to build more efficient teams of motors.
The effect also provides a useful way to examine how multiple motors cooperate with the dynamical formulations explored in this work.

To help understand this phenomenon, we take a simpler case of the same system, with the small modification of removing the thermal motion of the cargo (Fig.~\ref{fig:fasterthanfree}B).
By neglecting thermal fluctuations and taking the limit $\gamma \to 0$, the free variable of the cargo position $x$ is removed, since the cargo will quickly relax to the equilibrium position between the motors before their next jump, greatly simplifying analysis.
Now the state of the system can be completely described by the discrete distance between the two motors $\xi$.

By assuming a steady state distribution of the two motors and using the detailed balance constraint, we derived the velocity of the system by calculating the probability $P_\xi$ for the motors to be in state $\xi$ with the velocity contribution $v_\xi$, the full details of which are shown in SI\zref{app:two-motor-approximation}.
Assuming the backward rates are negligible for the most probable configurations (low $\xi$) allows us to solve for the velocities exactly~(SI\zref{app:harmonic-potential}).
The speedup in the D-AsEx model can then be understood by seeing that the probabilities $P_\xi$ decay exponentially in $k\Theta \xi^2$, while the velocities grow exponentially in $k\Theta$.
This allows for some states which, despite their low probability, contribute considerably to the mean velocity due to their exponential growth.
An analytic solution also allows us to trivially find the optimal $\Theta$ for a given $k$ in the D-AsEx model.
For physical values $k\sim 1$, $\Theta \approx 1/8$ maximizes the cooperativity, in good agreement with Fig~\ref{fig:fasterthanfree}, though the speedup is marginal at such small $k$ with $v^D/v_0 \approx 1.008$.
In general, we find that $\Theta \approx 1/8$ leads to optimal cooperation between two non-interacting motors in this limit, nearly independently of the $k$ value chosen.

We also find that in contrast to the D-AsEx model, the P-AsEx two-motor system monotonically decreases in $k\Theta$, where in the unlimited case, $\left<v^P\right> \approx e^{-\frac{k\Theta(1-\Theta)}{2}} \left<v^D\right>$.
Essentially, the decrease in the effective rate $w_0^P$ in the P-AsEx model for a given $\Theta$ always counters out the speed gains from the D-AsEx model.

It is important to note that a similar speedup effect has been observed in an analogous system modeling RecBCD enzymes, where two non-interacting motors are directly tethered to one another with a rate splitting that, at first glance, resembles the unlimited P-AsEx model~\cite{stukalin2005,stukalin2006}.
Their speedup is seemingly directly in contrast to our findings here, where the P-AsEx model forbids faster-than-unloaded
cooperation.
To resolve this apparent contradiction, we note that there are important differences between the work of~\citet{stukalin2005} and ours.

As noted before, their rate-splitting formulation is fundamentally different than the unlimited P-AsEx splitting in this work.
While we employ a scheme where the rates always receive the same $\Theta$ weighting on the change in energy, their weighting changes depending on if the jump is downhill or uphill in energy, essentially having $\Theta \rightarrow 1-\Theta$ when changing the sign of the energy difference for a jump.
When applying our P-AsEx splitting to their model, we find that the velocity similarly monotonically decreases with increasing energy cost, resolving the apparent contradiction.

It is still however worth exploring the other major difference between our works.
Their model uses a {\itshape linear} potential between states, rather than the harmonic potential used in this work.
However in~\citet{stukalin2005}, they consider only states where the motors are adjacent such that the harmonic and linear potentials can be mapped directly between one another, and still observe the same faster-than-unloaded effect.
A linear potential therefore cannot independently drive the disparity.
A more rigorous derivation of the mean velocity with a linear potential in the unlimited P-AsEx model shows that $\left<v\right> \approx 2\left(1+e^{\epsilon \Theta}\right)^{-1}$, where $\epsilon$ sets the energy scale of a single jump~(SI\zref{app:linear-potential}).

\subsection{Motor Number Scaling}
Motors often operate in teams much larger than two to move a burdensome cargo.
In many cases, the motors' locations along their tracks are sufficiently far apart to not significantly interact with one-another, such as in intraflagellar transport~\cite{prevo2015,pan2006,stepanek2016}.
However, in other cases, the motor heads along the track can significantly hinder one another~\cite{leduc2012,nishinari2005}.

Here, we examine the differences in motor number scaling behavior for these two cases under different models and parameters to see if there are meaningful qualitative differences in the scaling between the different models.
When considering the cooperativity of teams of motors, it is useful to measure the velocity of a team of motors normalized by the velocity of a {\itshape loaded} single motor $v_1$.
This gives a direct measure of how the velocity scales with the addition of extra motors, with ``perfect'' scaling being represented by $v_N/v_1 = N$.

Due to the large parameter space, we do several ``scans'' about a central point (Fig.~\ref{fig:no_excluded_volume}), varying the motor coil stiffness $k$, jump bias $\Delta \mu$, load-sharing factor $\Theta$, and the cargo's viscous drag coefficient $\gamma$ separately.
We, however, do not vary the base motor hopping rate $w_0$, as the scaling behavior is equivalent to varying the drag coefficient of the cargo when normalized by the single motor velocity -- i.e., for a given $w_0\gamma$, the normalized velocity $v(w_0\gamma)/v_0$ is fixed.

\subsubsection{Non-interacting Motors}
\begin{figure*}[hbt!]
  \begin{center}
    \includegraphics[]{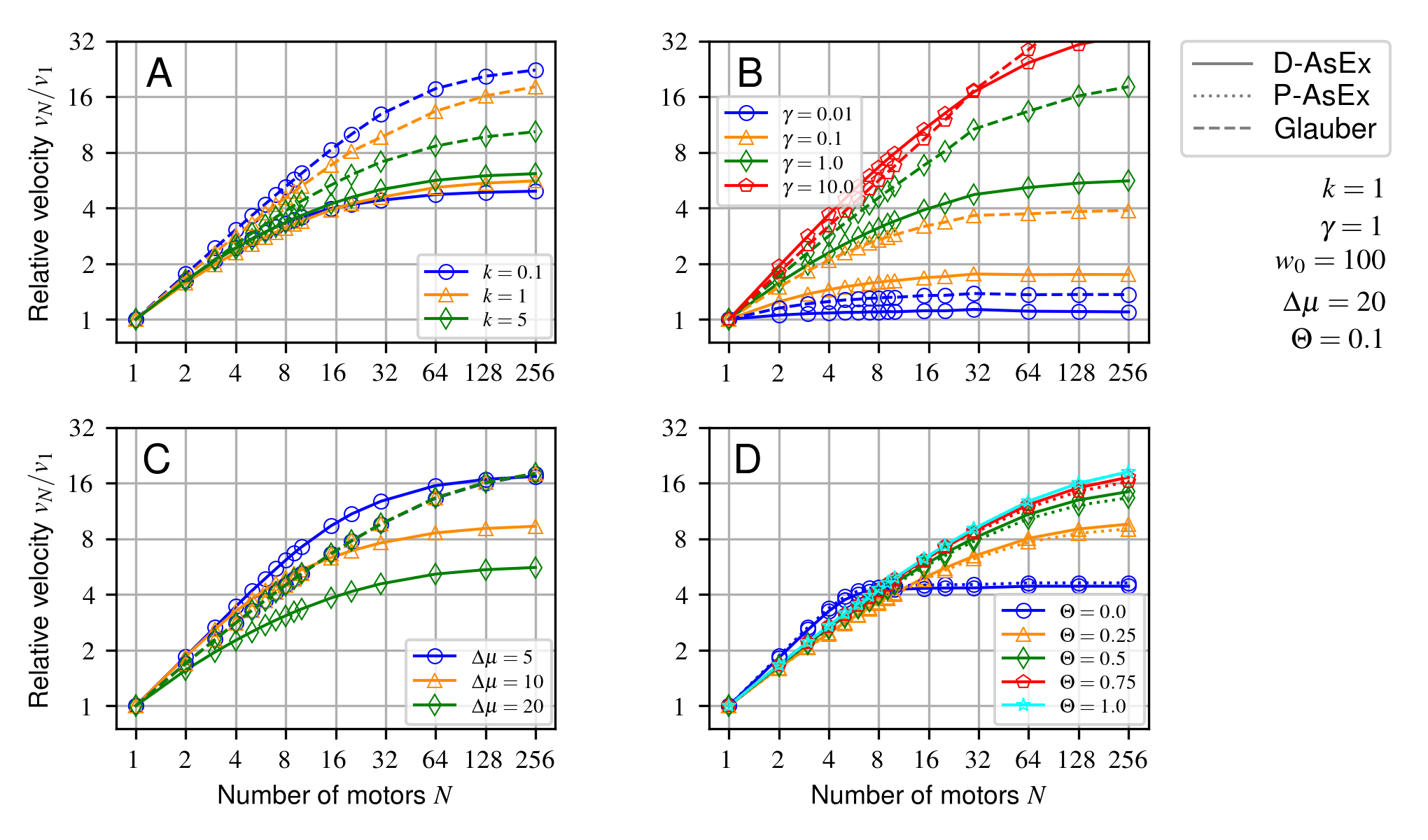}
    \caption{Relative velocities versus number of motors for various model parameters without lattice site exclusion.
      Parameters are varied in one dimension relative to the central point indicated in the figure.
      \textbf{\textit{(A)}} Varying motor-cargo coupling $k$,
      \textbf{\textit{(B)}} cargo damping $\gamma$.
      \textbf{\textit{(C)}} relative probability of backward stepping via $\Delta \mu$,
      \textbf{\textit{(D)}} and load-sharing factor $\Theta$.
    }\label{fig:no_excluded_volume}
\end{center}
\end{figure*}
The response of the system to changing $k$ is understandably complex, since it is the only parameter to be directly involved in both the motor hopping dynamics and the cargo motion dynamics~(Fig~\ref{fig:no_excluded_volume}A).
It is perhaps not surprising then that changing $k$ in different models can lead to qualitatively distinct results.
Increasing $k$ leads to a {\itshape decrease} in cooperativity with increasing motor number in the Glauber model, but an {\itshape increase} in cooperativity in the D-AsEx model.
The large scale separation of cooperativity in the Glauber model between the $k = 0.1$ and $k = 5$ is likely due to the large ceiling on the Glauber scaling potential relative to the D-AsEx model (SI\zref{app:param_scan}).
The Glauber single-motor velocities are very low compared to the unloaded velocities, with $v_1/v_0 \sim 0.034 - 0.04$, giving potential for a roughly 25-30 fold increase in velocity relative to the single-motor velocity.

\begin{figure*}[hbt!]
  \begin{center}
    \includegraphics[]{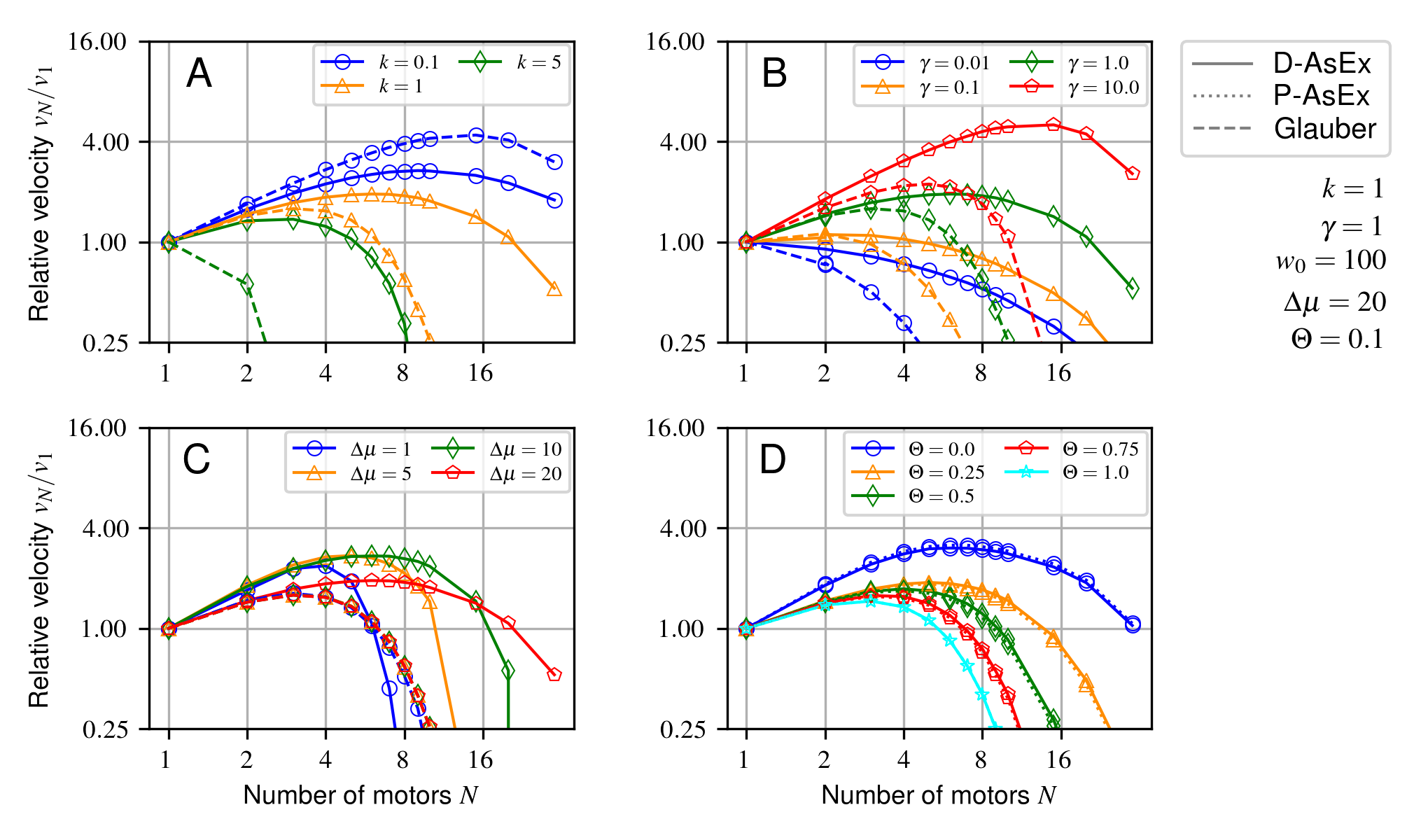}
    \caption{Relative velocities versus number of motors for various model parameters with excluded volume.
      Parameters are varied in one dimension relative to the central point indicated in the figure.
      \textbf{\textit{(A)}} Varying motor-cargo coupling $k$,
      \textbf{\textit{(B)}} cargo damping $\gamma$,
      \textbf{\textit{(C)}} relative probability of backward stepping via $\Delta \mu$,
      \textbf{\textit{(D)}} and load-sharing factor $\Theta$.
    }\label{fig:excluded_volume}
\end{center}
\end{figure*}

The cargo drag coefficient $\gamma$ also has a very strong effect on cooperativity, where increasing the drag increases the cooperativity, as shown in Fig.~\ref{fig:no_excluded_volume}B.
This is at first glance possibly counter-intuitive, since $\gamma$ increases the relaxation time and causes the cargo to lag behind the mean motor position for longer.
The cargo's backward position then in-turn lowers the overall average average forward jump rate, lowering the mean velocity.
However, the velocities are scaled relative to the single-motor velocities, which can also be drastically diminished by increased cargo drag (SI\zref{app:param_scan}), counteracting the relaxation time cost.
The velocities also saturate at large motor numbers, though the saturation point is strongly parameter dependent.
The full cargo relaxation time decreases inversely proportional to the number of motors in the train $\gamma/(N\cdot k)$, such that, roughly, when $\gamma/(N\cdot k) \ll w_0^{-1}$, the velocity saturates.
While there is a clear distinction in the behavior between the two rates when varying cargo drag, their behavior is similar enough to make it difficult to experimentally distinguish between the two rate formulations.
It is still useful to note that high-drag environments could be useful to study cooperativity since they can provide much larger changes in relative velocity over much larger motor numbers.

The final way to directly adjust a transport timescale, as with $k$ and $\gamma$, is by adjusting the forward jump bias $\Delta \mu$, or equivalently the base backward jumping rate $w_0 e^{-\beta \mu}$, as shown in Fig.~\ref{fig:no_excluded_volume}C.
The motor number scaling behavior of the D-AsEx model is inverted with respect to $\Delta \mu$: {\itshape increasing} $\Delta \mu$ causes a {\itshape decrease} in cooperativity.
As with the case for $\gamma$, this inversion is due to the single motor velocity of the high $\Delta \mu$ case being much closer to the saturation velocity $\ v_0$ than the low $\Delta \mu$ case.
The Glauber model, however, shows no virtually no change in behavior with changing $\Delta \mu$.
This is due to the near unity velocity ratio for a given motor-cargo configuration 
\begin{equation}
  \frac{w_f(\delta x)-w_b(\delta x,\Delta\mu_1)}
  {w_f(\delta x)-w_b(\delta x,\Delta\mu_2)} \approx 1
\end{equation}
for any combination $(\Delta\mu_1, \Delta\mu_2) \gtrapprox 3$ in the Glauber model within our reference set of parameters.
This relation does not hold well for the AsEx models, however, with the ratio potentially diverging and becoming negative with increasing $\delta x$.
The nearly complete insensitivity of the Glauber rates to $\Delta \mu$ could stand allow for a simple experimental test to distinguish between the Glauber and AsEx formulations of motion.
By adjusting the ATP concentration on a system such as a microtubule motility assay, there should be no change in relative velocities with increasing motor number of microtubules if the Glauber rates are appropriate when controlling for microtubule drag.

It is clear that adjusting the timescales, forces, and energies involved in transport can dramatically effect the velocities of cargo transport.
It is also clear from this work that the model choice can be equally, if not more important in deciding the net transport velocity given these known properties of a system, so we further observe the effects of varying the main model parameter $\Theta$ on transport.
We find that for the chosen parameters, the scaling behavior is almost independent of $\Theta$, with the exception of $\Theta = 0$, which saturates at low motor numbers due to a high single-motor velocity (Fig.~\ref{fig:no_excluded_volume}D).
This suggests that a rapid change in scaling with motor number could be indicative of an AsEx model with small $\Theta$.
We also find that there is virtually no difference between the D-AsEx model and the P-AsEx model for this set of parameters.
As shown earlier, the leading term in the D-AsEx and P-AsEx models can be mapped to between one another via a change in the effective rate constant  $w^\mr{P}_0=w^\mr{D}_0e^{-\frac{k\Theta}{2}(1-\Theta)}$.
For $k = 1$, this means that in the worst case of $\Theta = 0.5$,  $w_0^P \approx 0.78~w_0^D$, which is far from the order of magnitude changes needed to change the scaling behavior significantly, resulting in nearly identical results between the two models.
In the higher $k$ regime, however, the P- and D-AsEx results should significantly split.
For example, if $k = 20$, the effective rate in the P-AsEx model for $\Theta = 0.5$ will be $w_0^P \approx 0.08~w_0^D$, which will give a factor of $\sim 2-3$ difference in velocities between the two models, similar to the $\gamma$ case.

\subsubsection{Interacting Motors}
The monotonic increase in cargo velocity with increasing motors hinges on the motors' ability to cooperate in cargo transport without colliding or hindering each other through direct motor-motor interactions.
This is a reasonable assumption when motors are attached to the cargo at a sufficiently large distance from each other, as in the case of intraflagellar transport~\cite{prevo2015}.
In other cases, collisions of molecular motors and traffic congestion effects may be an important factor, since the diameter of a microtubule is merely 25 nm, barely above the size of a kinesin motor~\cite{lodish2000}.
For this reason we also examine the effect of motor head excluded volume interactions on transport for each of our rate formulations.
To do this, we assume that the motor heads are not allowed to occupy the same lattice position, and that they are all attached at the some point to some cargo.
This is likely an overly simplistic assumption, since multiple motors will not be able to attach to the same site on the cargo, but a full exploration of the effects of this positioning is beyond the scope of this work.
This limit is still however useful to show the large differences in model behaviors when motor collisions are included.

Possibly the most obvious parameter to have a strong impact on dynamics involving excluded volume interactions is the coil stiffness $k$.
Since the leading motor will maintain an average distance to the vesicle of $\langle n-x\rangle \geq N/2$, a large $k$, combined with a large number of motors $N$, will stall the front motor and therefore the whole transport process.
We see this effect clearly in both the Glauber and D-AsEx rate formulations in Fig.~\ref{fig:excluded_volume}A, where for high $k$ the motor hindrance rapidly stalls cargo motion, with the Glauber rates at $k=5$ showing no speedup for even two motors.
This could indicate a reason why typical coil stiffnesses are on the order of one \cite{kawaguchi2001,lansky2015}, as the lower stiffnesses allow for more cooperative motion between the motors, while not requiring large extensions to generate force.

Interestingly, for the $k$ scan in Fig.~\ref{fig:no_excluded_volume}A, while the Glauber models uniformly showed higher cooperativity than the D-AsEx model in the non-interacting case, the introduction of motor exclusion causes the Glauber model to perform unfavorably in the high $k$ limit compared to any variant of the D-AsEx model.
From the definition of the D-AsEx forward rate (Eq.~\ref{eq:d-asex-forward}), it is clear that the forward rate decays exponentially with a length scale $\ell_D \sim (\Theta k)^{-1}$, compared to that of the Glauber rates $\ell_G \sim k^{-1}$ (Eq.~\ref{eq:glauber-forward}).
This slow decay clearly favors cooperativity in the jammed environment examined here, since the lead motor will have a large extension $\delta x \sim N/2$.

As with the non-interacting case, the cargo drag coefficient $\gamma$ also exhibits a very strong effect monotonic increase of cooperativity with increasing $\gamma$ ( Fig.~\ref{fig:no_excluded_volume}B). 
In contrast with the non-interacting case, in the low $\gamma$ limit, motors are always anti-cooperative for both the Glauber and D-AsEx models.
Also unlike the non-interacting case, in crowded environments there is a strong distinction between the two models.
The Glauber rates are much more sensitive to molecular crowding, causing catastrophic collapse in cooperativity at motor numbers much lower than their AsEx counterparts at all observed drag coefficients.

The effect of the forward jump bias $\Delta \mu$ with motor head exclusion is much more ambiguous than in the non-interacting case (Fig.~\ref{fig:excluded_volume}C).
With the D-AsEx rates at low motor numbers, $\Delta \mu = 5$ and $\Delta \mu = 10$ show virtually identical scaling, with $\Delta \mu = 20$ underperforming.
At high motor numbers, the behavior crosses over, with the high $\Delta \mu$ outperforming the low $\Delta \mu$ cases, with the $\Delta \mu = 5$ case rapidly becoming anti-cooperative at intermediate motor number.
However, as with the non-interacting case, the cooperativity in the Glauber model is almost completely insensitive to $\Delta \mu$, with almost all values simulated being indistinguishable.

Unlike $\Delta \mu$, the value $\Theta$ shows a completely unambiguous decrease in cooperativity with increasing $\Theta$ (Fig.~\ref{fig:excluded_volume}).
This can be understood by the same argument for the collapse in the cooperativity using Glauber rates with high $k$.
The stalling length for the forward rates $\ell \sim (k\Theta)^{-1}$ increases with $\Theta$, allowing for less stalling of the lead motor.
This of course can not go on forever, the backward rates, which grow as $\sim e^{-\Delta \mu + k(1-\Theta)\delta x}$, eventually will overcome the forward rates, leading to anti-cooperativity for large motor numbers. 

\section{CONCLUDING REMARKS}
There are clear and distinct differences in the physical behavior of motor-cargo systems depending on the underlying rate formulation chosen.
These choices can allow for both novel effects, such as the faster-than-unloaded behavior shown in the two-motor D-AsEx system at low viscosity, drastic qualitative and quantitative shifts in the force-velocity relationship for a single motor under load, and similarly drastic changes in the cooperativity of large numbers of motors working in teams.

Due to the near equivalence of the D- and P-AsEx models at small and fixed $k$, experiments that differentiate between these two possible model choices require control of the coil stiffness $k$ over roughly an order of magnitude, though with a potentially large cooperativity payoff.
The coil stiffness could be directly altered in purely engineered systems, or potentially in biological systems by mutations, direct substitutions, or interactions with external molecules.
The Glauber and AsEx rate formulations, however, are starkly distinct in ways which are more easily directly observed.
The most notable difference between the two models is in their sensitivity to $\Delta \mu$, which can be adjusted by changing available chemical fuel.
The Glauber model shows no sensitivity to $\Delta \mu$ in the relative velocity $v_N/v_1$ under the observed parameters, while the AsEx model shows a clear change in scaling behavior with $\Delta \mu$, regardless of motor collisions (Figs.~\ref{fig:no_excluded_volume}C,\ref{fig:excluded_volume}C), with similar insensitivity in the single-motor behavior (SI\zref{app:param_scan}).

These three models are also not the only possible models that fit the necessary and relatively weak detailed balance constraint.
Further work could place constraints on which classes of models are physical and what the more general properties of these models are.
For example, it is clear that the Glauber and P-AsEx models cannot, for the potentials explored, exhibit the faster-than-unloaded behavior that is observed in the model employed by Stukalin and the D-AsEx model employed here.
The speedup is potentially a general feature of some class of models which should be explored in greater detail.

\section{ACKNOWLEDGEMENTS}
A.-.S.S and R.B. thank the ERC grant MembranesAct StG 337283 and the Alexander von Humboldt Foundation for the postdoctoral fellowship provided for R.B.
Special thanks to Udo Seifert for useful discussions and Daniel Schmidt for an early oversight for the project.

\bibliography{paper.bib}

\end{document}


\section{Low-viscosity two-motor analytic approximation}\zlabel{app:two-motor-approximation}
We let $P_\xi$ be the unnormalized probability for the two motor heads to be at a distance of $\xi$ from each other, and assume a steady state distribution so that $\frac{\mathrm{d}P_\xi}{\mathrm{d}t}=0 \, \, \forall~\xi$.
Starting with some unknown $P_0$ we can iteratively calculate $P_{n+1}$ as a function of all $P_{m\leq n}$ using the detailed balance constraints,
\begin{subequations}
\begin{equation}
  P_1 = 2 P_0\frac{w_\mr{f}(0) + w_\mr{b}(0)}
  {w_\mr{f}(-\frac{1}{2}) + w_\mr{b}(\frac{1}{2})}
\end{equation}
\begin{equation}
  P_2 = P_1 \frac{w_\mr{f}\left(\frac{1}{2}\right) + w_\mr{f}\left(-\frac{1}{2}\right) +
    w_\mr{b}\left(\frac{1}{2}\right) + w_\mr{b}\left(-\frac{1}{2}\right)}
  {w_\mr{f}\left(-1\right)+w_\mr{b}\left(1\right)}
  - 2 P_0 \frac{w_\mr{f}\left(0\right) + w_\mr{b}\left(0\right)}
  {w_\mr{f}\left(-1\right)+w_\mr{b}\left(1\right)}
\end{equation}
\begin{equation}
  P_{\xi+1} = 
  P_\xi\frac{w_\mr{f}\left(\frac{\xi}{2}\right) + w_\mr{f}\left(-\frac{\xi}{2}\right) +
    w_\mr{b}\left(\frac{\xi}{2}\right) + w_\mr{b}\left(-\frac{\xi}{2}\right)}
  {w_\mr{f}\left(-\frac{\xi+1}{2}\right) + w_\mr{b}\left(\frac{\xi+1}{2}\right)}
  -P_{\xi-1}\cdot \frac{w_\mr{f}
    \left(\frac{\xi-1}{2}\right) + w_\mr{b}\left(-\frac{\xi-1}{2}\right)}
  {w_\mr{f}\left(-\frac{\xi+1}{2}\right)+w_\mr{b}\left(\frac{\xi+1}{2}\right)},
\end{equation}
\end{subequations}
which can be solved exactly in terms of $P_0$
\begin{equation}\zlabel{eq:probabilities}
  P_{\xi>0} = 2 P_0 \prod_{i=0}^{\xi-1}
  \frac{w_f\left(\frac{i}{2}\right) +w_b\left(-\frac{i}{2}\right)}
          {w_f\left(-\frac{i+1}{2}\right) +w_b\left(\frac{i+1}{2}\right)}.
\end{equation}
Similarly, the velocity of the cargo depending on the current motor configuration can be calculated from the jump rates as
\begin{equation}
  v_\xi = \frac{1}{2}\left(
    w_\mr{f}\left(\frac{\xi}{2}\right) + w_\mr{f}\left(-\frac{\xi}{2}\right) -
    w_\mr{b}\left(\frac{\xi}{2}\right) - w_\mr{b}\left(-\frac{\xi}{2}\right)
  \right),
\end{equation}
and the long-term velocity under an equilibrium motor configuration is then obtained by summing over all motor configurations weighted by their respective probabilities
\begin{equation}\zlabel{eq:velocity}
  \langle v \rangle=\frac{\sum_{\xi=0}^{\infty} v_\xi\cdot P_\xi}{\sum_{\xi=0}^{\infty} P_\xi}=
  \sum_{\xi=0}^{\infty} v_\xi\cdot p_\xi,
\end{equation}
where $p_\xi$ is the unit normalized probability such that $\sum_{\xi=0}^\infty p_\xi = 1$.

\section{Two-motor velocity with harmonic potential}\zlabel{app:harmonic-potential}
When considering motors with a low relative backward hopping probability ($\Delta mu \gg 1$) and vanishing drag in the unlimited variants of the AsEx models, the velocity can be solved exactly.
In both of the AsEx models, setting the $w_b$ terms to zero and using the derivation outlined above, it is trivial to show that the normalized probabilities
\begin{equation}
  p_\xi = \frac{(2-\delta_{\xi,0})e^{-\frac{k\Theta}{2} \xi^2}}{\sum_{\xi=0}^\infty (2-\delta_{\xi,0})e^{-\frac{k\Theta}{2} \xi^2}},
\end{equation}
where the sum in the denominator is an elliptic theta special function and can be written $\mathcal{\theta}_3(0,e^{-\frac{k\Theta}{2}})$.

The velocity contribution of a given configuration $v_\xi$ in this limit for the P-AsEx model can be simply written
\begin{equation}
  v_\xi^{\mr P} = w_0 e^{-\frac{k\Theta}{2}}\cosh\left(\frac{k\Theta \xi}{2}\right).
\end{equation}
Then, putting the velocity and probabilities together, we arrive at the mean velocity in the low $\gamma$ and high $\Delta \mu$ limit
\begin{align}
  \left<v^P\right> &= \frac{w_0 e^{-\frac{k\Theta}{2}}}{\mathcal{\theta}(0,e^{-\frac{k\Theta}{2}})}
             \sum_{\xi=0}^\infty{(2-\delta_{\xi,0})e^{-\frac{k\Theta}{2} \xi^2}}\cosh\left(\frac{k\Theta\xi}{2}\right)\notag\\
           &= w_0e^{-\frac{3}{8}k\Theta}\frac{\mathcal{\theta}_2(0,e^{-\frac{k\Theta}{2}})}
             {\mathcal{\theta}_3(0,e^{-\frac{k\Theta}{2}})}.
\end{align}
This function monotonically decreases in $k\Theta$, and should act as a maximum velocity for a two-motor system, showing that the P-AsEx model as formulated cannot result in a two-motor speedup relative to the unloaded motors.

The D-AsEx velocity can similarly be calculated using the generalized velocity derivation outlined above.
However, since the two unlimited AsEx models can be mapped between one another, we can simply write
\begin{equation}
  \left<v^D\right> = e^{\frac{k\Theta(1-\Theta)}{2}} \left<v^P\right> = w_0e^{\frac{1}{2}k\Theta(\frac{1}{4}-\Theta)}\frac{\mathcal{\theta}_2(0,e^{-\frac{k\Theta}{2}})}
             {\mathcal{\theta}_3(0,e^{-\frac{k\Theta}{2}})}
\end{equation}
Noting that for $k\Theta \lessapprox 4$ the ratio of the elliptic theta functions is $\approx 1$, we can approximate
\begin{equation}
  \left<v^D\right> \approx w_0 e^{\frac{1}{2}k\Theta(\frac{1}{4}-\Theta)} 
\end{equation}
for small $k\Theta$.
This clearly allows for speedups for any $\Theta < 1/4$ with the optimal $\Theta = 1/8$.

\section{Two-motor velocity with linear potential}\zlabel{app:linear-potential}
Since the prior work on modeling the RecBCD protein observed a two-motor speedup used a linear potential between the motor heads, it is useful to explore if this is possible using a linear potential but with a cargo intermediary in the unlimited P-AsEx formulation.
The potential can be simply written $V(\delta x) = \epsilon \left|\delta x\right|$, where $\epsilon$ sets the energy scale of a single jump.
The forward rate is then $w_f(\delta x) = w_0 e^{-\epsilon \Theta (\left|\delta x + 1\right| - \left|\delta x\right|)}$.
Noting that the forward rate can take on one of three values since the displacement can only take integer and half-integer values
\begin{subequations}
  \begin{equation}
    w_f(\delta x \leq -1) = w_0 e^{\epsilon \Theta}\\
  \end{equation}
  \begin{equation}
    w_f(\delta x = -\frac{1}{2}) = w_0\\
  \end{equation}
  \begin{equation}
    w_f(\delta x \geq 0) = w_0e^{-\epsilon \Theta},
  \end{equation}
\end{subequations}
and setting the backward rates $w_b = 0$, the mean velocity can be simply calculated using the relations from Equations SI\zref{eq:probabilities} and SI\zref{eq:velocity}
\begin{equation}
  \left<v\right> = \frac{e^{-\epsilon \Theta}\left(1 + 2\sinh(\epsilon\Theta)\right)}{1+\sinh(\epsilon\Theta)} \approx \frac{2}{1+e^{\epsilon \Theta}}.
\end{equation}
As with the harmonic potential in the unlimited P-AsEx model, this monotonically decreases in $\epsilon\Theta$.
\clearpage

\section{Single motor response in reference set}\zlabel{app:param_scan}
\begin{figure}[hbt!]\nonumber
  \centering
  \includegraphics[]{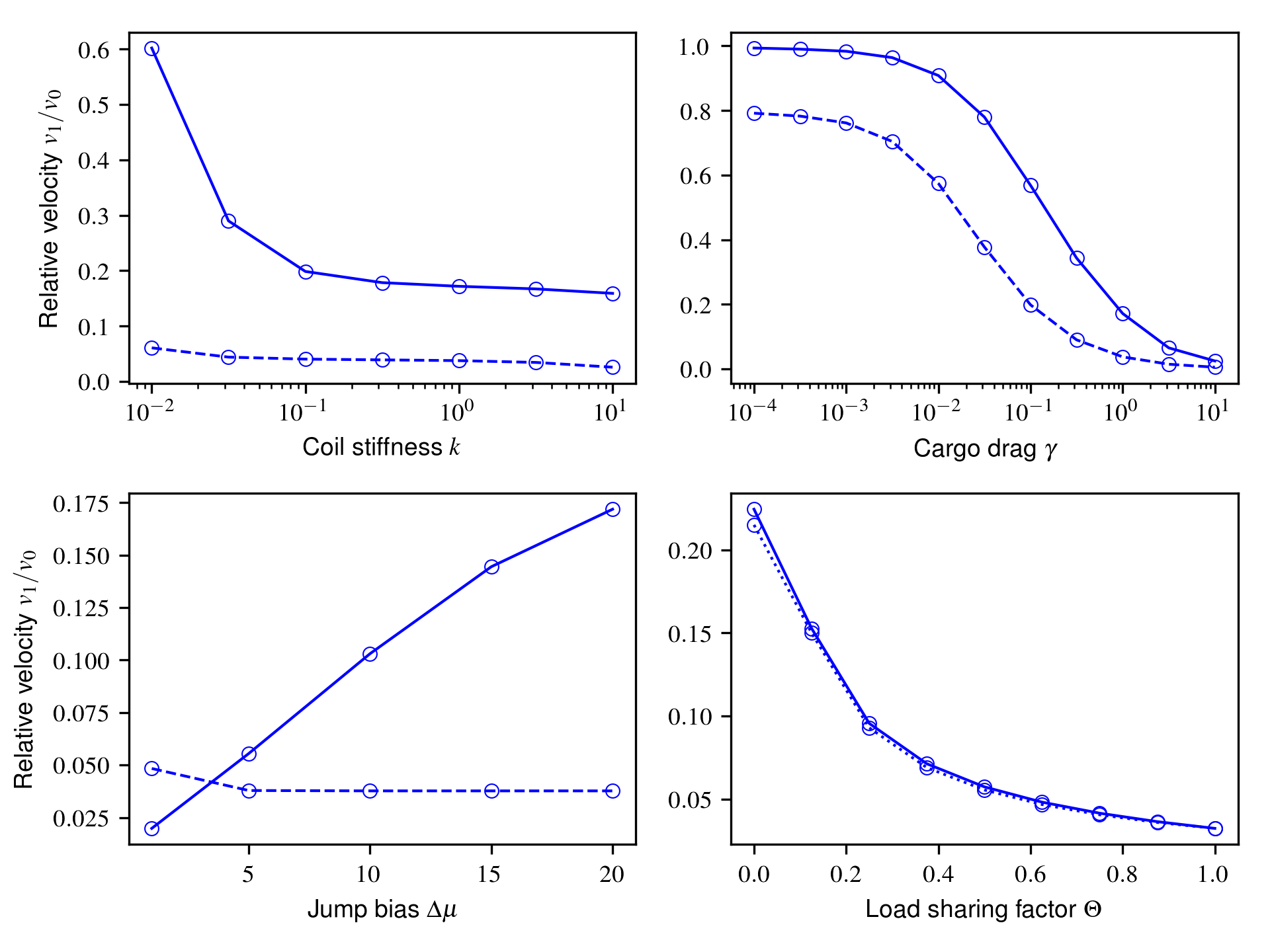}
  \caption*{Relative velocities $v_1/v_0$ for single motors about the central point $\gamma=1$, $\Delta\mu=20$, and $w_0=100$, $k=1$.
    D-AsEx shown in solid lines, P-AsEx in dotted lines, and Glauber in dashed lines.
    Scan in 
    \textbf{\textit{(A)}} coil stiffness $k$,
    \textbf{\textit{(B)}} cargo drag $\gamma$,
    \textbf{\textit{(C)}} jump bias $\Delta \mu$,
    \textbf{\textit{(D)}} and load sharing factor $\Theta$.
  } 
\end{figure}
\clearpage

\section{Two-motor velocities in P-AsEx formulation}\zlabel{app:p-asex}
\begin{figure}[hbt!]\nonumber
  \centering
  \includegraphics[]{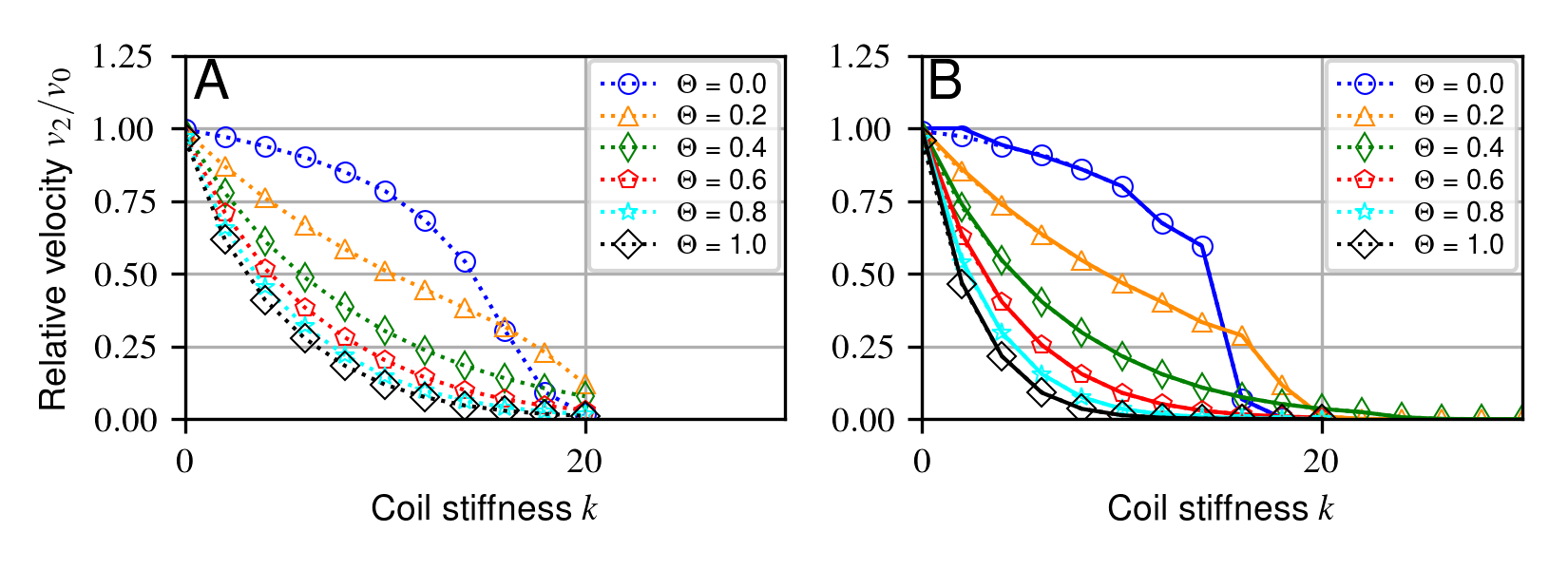}
  \caption*{Relative velocities $v_2/v_0$ for two non-excluding motors with parameters in our dimensionless units $\gamma=10^{-3}$,$\Delta\mu=20$, and $w_0=60$.
    \textbf{\textit{(A)}} P-AsEx rates simulation for two non-interacting motors.
    \textbf{\textit{(B)}} P-AsEx rates simulation (dotted line) vs analytic approximation (full line) with no thermal motion of the cargo.
  } 
\end{figure}